\documentstyle[12pt,twoside,fleqn,epsf]{article}

\begin{document}

\begin{center}
{\hfill {\normalsize KUNS-1561}}\\
{\large\bf Delayed Collapse of Protoneutron Stars\\
with Kaon Condensate}\\
Masatomi Yasuhira$^a$ and Toshitaka Tatsumi$^b$\\
Department of Physics, Kyoto University, JAPAN\\
{\small $^a$ yasuhira@ruby.scphys.kyoto-u.ac.jp \\
$^b$ tatsumi@ruby.scphys.kyoto-u.ac.jp}
\end{center}

\begin{abstract}
Equation of state with kaon condensate is derived
for isentropic and neutrino-trapped matter.
Both are important ingredients to study the delayed collapse of 
protoneutron stars. 
Solving the TOV equation,
we discuss the static properties of protoneutron stars
and implications for their delayed collapse.
\end{abstract}

\section{Introduction}

After supernova explosions,
protoneutron stars (PNS)
are formed
with hot, dense and neutrino-trapped matter.
They usually evolve to cold ($T\simeq0$) neutron stars
through the deleptonization and cooling stages.
But some of them may collapse to low-mass black holes
during these stages by softening the equation of state (EOS)
due to the hadronic phase transitions\cite{BrownBethe}.
This is called the delayed collapse;
as a typical example
neutrinos from SN1987A were observed at Kaomiokande,
but no pulsar yet,
which suggests the possibility of the delayed collapse of
SN1987A.

Kaon condensation,
one of the candidates of the hadronic phase transitions,
has been studied by many authors
mainly at zero temperature
since first suggested by Kaplan and Nelson\cite{Kaplan}.
We know that the kaon condensation gives rise to
the large softening of EOS.
Recently there appear a few works
about kaon condensation at finite temperature,
but there was no consistent theory based on chiral symmetry.
We have presented a new formulation to treat fluctuations
around the condensate based on chiral 
symmetry\cite{TTMY98l}\cite{TTMY}.

Here we show the essential idea of our formulation
and then discuss the properties of PNS and the possibility
of the delayed collapse.

\section{Formulation}

We use the nonlinear sigma model ( as a chiral Lagrangian )
and treat fluctuation fields around the condensate 
by making full use of the idea of chiral rotation.
Chiral Lagrangian on the chiral manifold
$G/H \simeq SU(3);$ $G=SU(3)_L\times SU(3)_R$, $H=SU(3)_V$
is specified by the unitary matrix
$U(\phi_a)=\exp(2iT_a \phi_a/f)$
with Goldstone fields:$\phi_a$,
generators of $SU(3)$ Lie algebra:$T_a$ and
pion decay constant:$f$.

Fluctuation around the condensate
is written as $U(\langle K^\pm \rangle, \tilde\phi_a)$
$=$
$\zeta\eta^2\zeta$.
$\eta$ corresponds to a chirally rotated state
from the meson vacuum:
$\eta$
$=$
$\exp\{i(V_+ K^- + V_- K^+)/\sqrt{2}f\}$
with $V_\pm=T_4\pm iT_5$ and
$K^\pm=(\phi_4\pm i\phi_5)/\sqrt{2}$.
On the other hand,
the classical condensate $\zeta$ is also chirally rotated state:
$\zeta$
$=$
$\exp\{i(V_+\langle K^- \rangle + V_- \langle K^+\rangle)/\sqrt{2}f\}$.
The most important field we are interested in
is the fluctuation field around the condensate.
In the usual approach
the meson fields are separated directly:
$\phi_a = \langle \phi_a \rangle+\tilde{\phi}_a$\cite{TandE}.
After putting this form into the Lagrangian,
we find
the integration measure is complicated and
the Lee-Yang term appears in the path integral.
We have therefore used another separation
by the use of the successive chiral rotations;
This can be regarded as a separation of zero-mode
or setting of the local coordinates around the condensed point.
Thus the Lee-Yang term disappears
and we can take the flat curvature to one-loop order\cite{TTMY}.


With this idea, we have performed the imaginary-time path integral to
one-loop order and then derived the thermodynamic functions in a
transparent way\cite{TTMY}. The $KN$ interactions have been treated
self-consistently within the Hartree approximation, and consequently
the thermodynamic potential consists of an infinite series of many
one-loop diagrams. There appear divergent integrals, but we can
renormalize them properly. Then the dispersion relations for kaonic
excitations, which are fundamental objects to study the thermal
properties of the condensed phase, are derived. We have two modes in
the condensed phase: a
Goldstone-like soft mode and a very massive mode, which correspond to
$K^-$- and $K^+$-mesonic excitations, respectively. Hence, the thermal
loops due to the soft mode play an important role in the thermodynamic 
functions.

\section{Numerical Results and Discussions}

With thermodynamic potenital,
we can study the nature of kaon condensed state
at finite temperature
and then discuss some implications 
on the delayed collapse of PNS.
We, hereafter, use the heavy-baryon limit for nucleons\cite{TTMY98l}.
We show the phase diagram, EOS
and then discuss the properties of PNS
where thermal and neutrino-trapped effects
are very important.

\begin{figure}[htb]
 \begin{minipage}{0.49\textwidth}
  \epsfsize=0.95\textwidth
  \epsffile{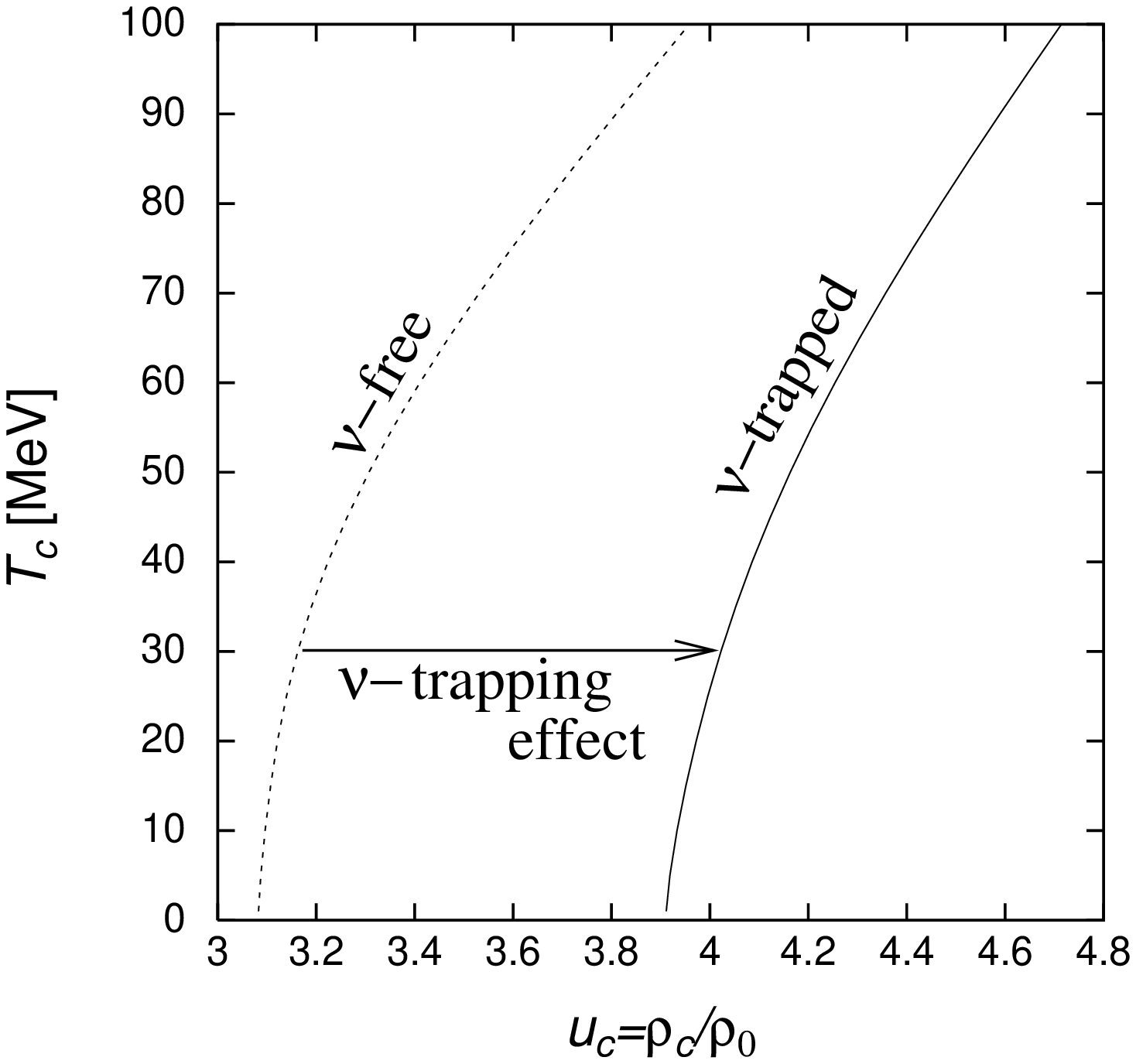}
\caption{
Phase diagram:
$\nu$-trapped($Y_{le}$
$=$
$0.4$) [solid line] 
and $\nu$-free cases [dashed line]. }
  \label{fig:pd}
 \end{minipage}%
 \hfill~%
 \begin{minipage}{0.49\textwidth}
  \epsfsize=0.95\textwidth
  \epsffile{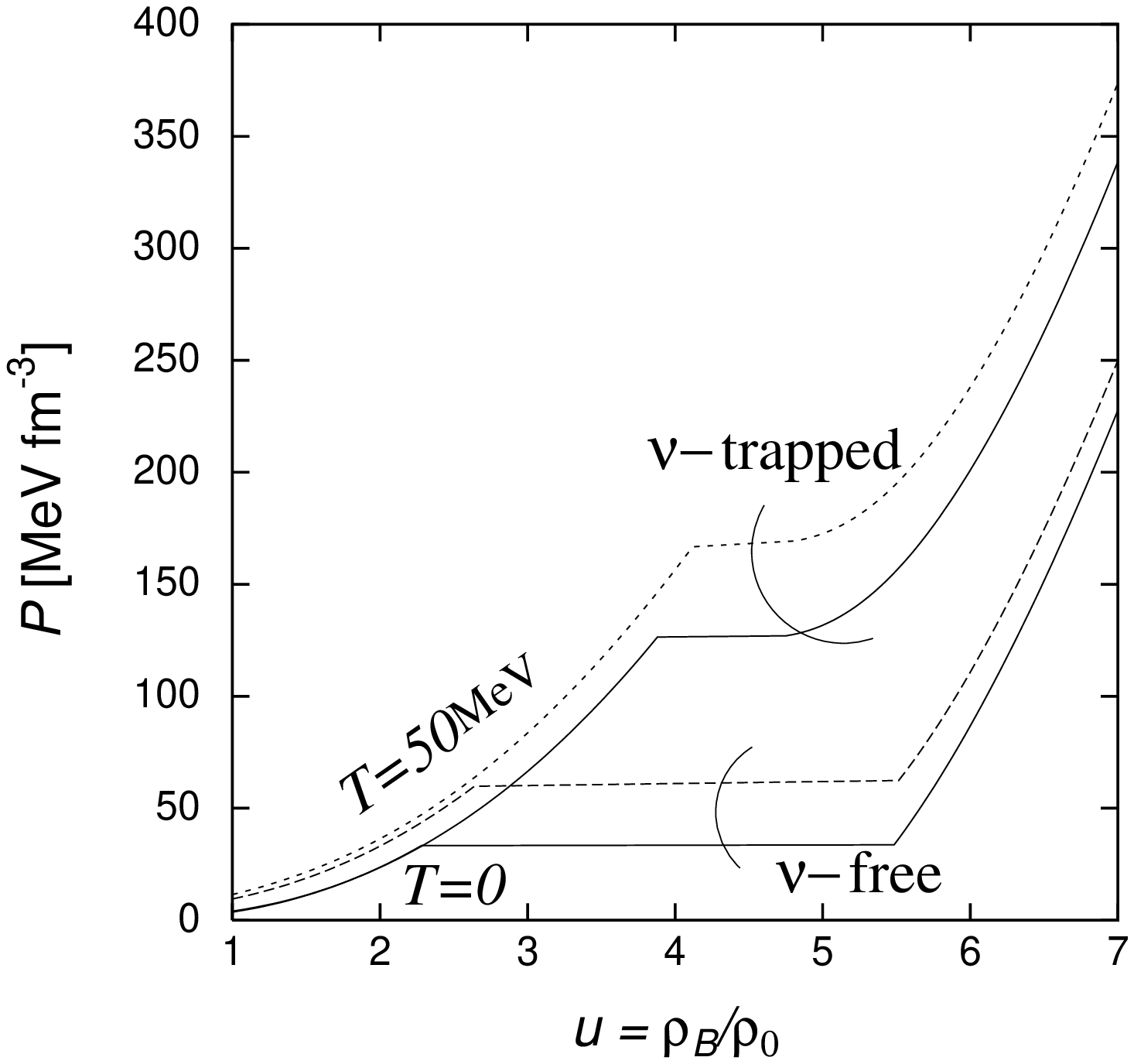}
\caption{
EOS for isothermal case: $T=0$ [solid line] 
and $T=50$MeV [dashed line]. 
}
  \label{fig:press}
 \end{minipage}%
\end{figure}

First we show the phase diagram 
and the isothermal EOS
in Figs.\ref{fig:pd},\ref{fig:press}.
In the neutrino-trapped case
we set $Y_{le}=Y_e+Y_{\nu_e}=0.4$ where
$Y_e$($Y_{\nu_e}$) is the electron(electron-neutrino)
number per baryon,
while $Y_{\nu_e}=0$ in the neutrino-free case.
Both of the thermal and neutrino-trapped effects
largely
suppress the occurence of kaon condensation.
The reason for the latter case
may be understood from
the threshold condition
for kaon chemical potential,
$\mu_K=\mu_e-\mu_{\nu_e}$:
$\mu_{\nu_e}>0$ in the neutrino-trapped case
while $\mu_{\nu_e}=0$ in the neutrino-free case,
which means kaons should wait to condense until
its energy further decreases.
Both effects also stiffen the EOS in the condensed phase
as well as the normal phase.
These effects are more pronounced in the condensed state
(see Fig.\ref{fig:press}).
Kaon condensation is the first order phase transition
and thereby
EOS includes thermodynamically unstable region.
We applied the Maxwell construction to obtain the equilibrium curve
for simplicity
though, restrictly speaking, we need to take the
Gibbs condition\cite{gle}.

\begin{figure}[htb]
 \vspace{2mm}
 \begin{minipage}{0.49\textwidth}
  \epsfsize=0.95\textwidth
  \epsffile{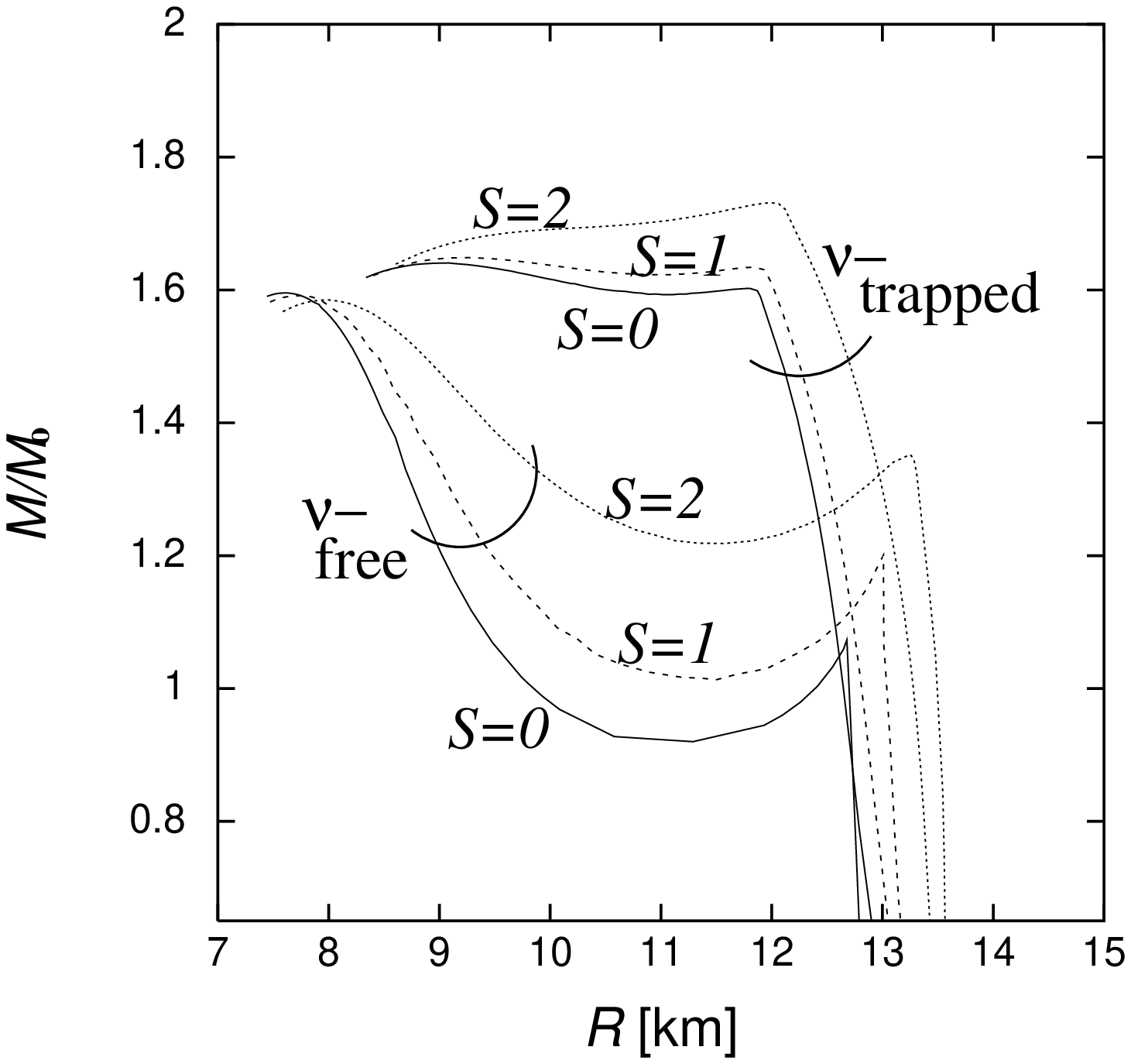}
 \caption{Mass-radius curve for PNS in $\nu$-trapped($Y_{le}=0.4$)
and $\nu$-free cases.
}
 \label{fig:MR}
 \end{minipage}%
 \hfill~%
 \begin{minipage}{0.49\textwidth}
  \epsfsize=0.95\textwidth
  \epsffile{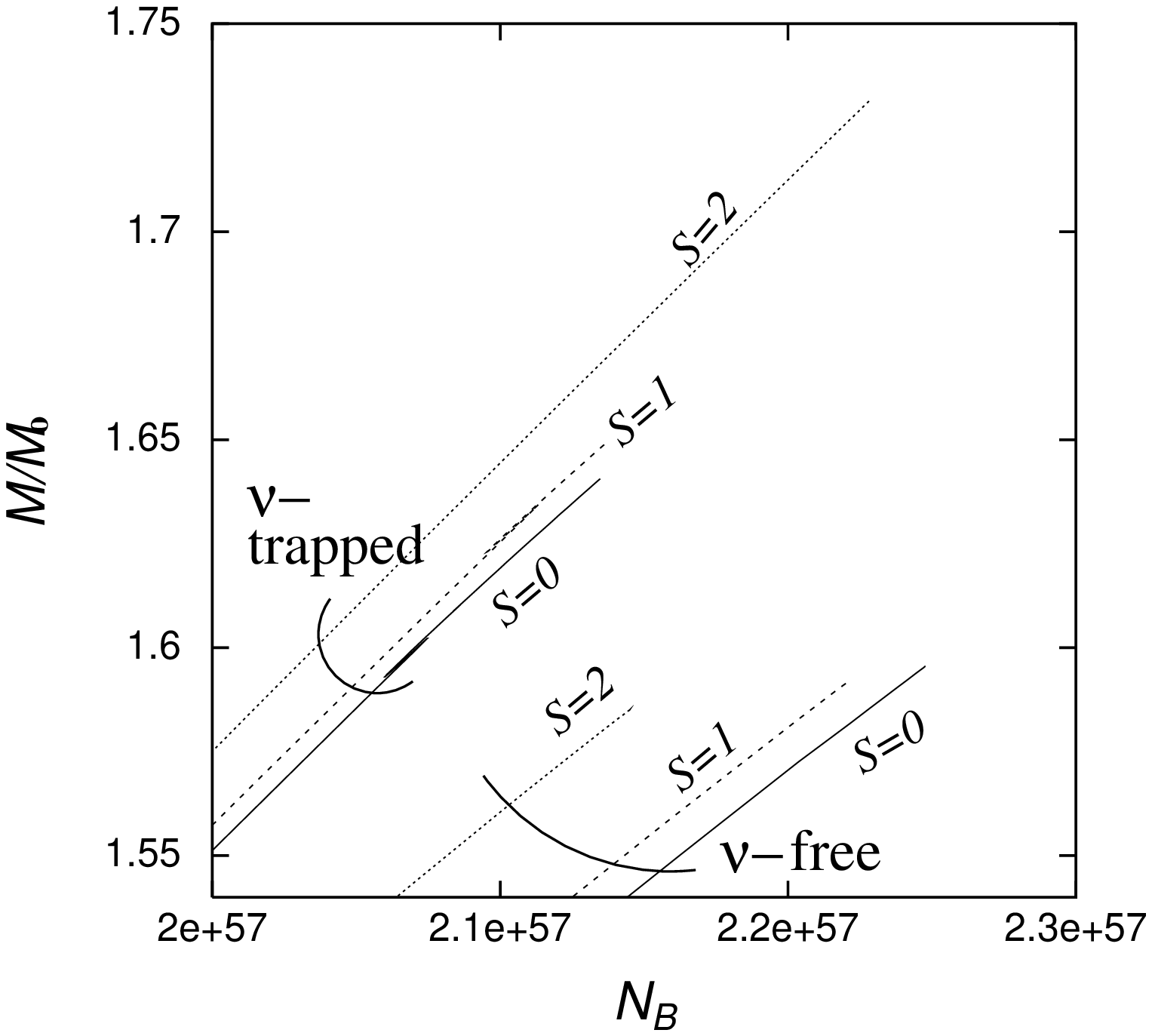}
 \caption{Total baryon number versus mass for PNS.
}
 \label{fig:NBM}
 \end{minipage}%
\end{figure}

Solving the TOV equation with the EOS,
we can study the nature of PNS for which
the isentropic situation is relevant.
In Fig.\ref{fig:MR} we show
the mass-radius relation for the neutrino-trapped and -free cases
with entropy per baryon $S=0,1$ or $2$.
Once kaon condensation occurs in the core of the star,
radius becomes smaller 
and
gravitationaly unstable branch appears
because of the large softening of EOS.
Both of the thermal and neutrino-trapped effects
make the radius larger.

To discuss the possibility of delayed collapse of PNS,
the total baryon number $N_B$ should be fixed
as a conserved quantity during the evolution.
In Fig.\ref{fig:NBM}
we show the mass versus total baryon number for
gravitationaly stable PNS.
Each terminal point represents maximum mass and maximum total baryon number
in each configuration.
If initial mass exceeds the terminal point,
the star should collapse into a black hole (not a delayed collapse but an
usual formation of a black hole). We have shown the neutrino-trapped
and -free cases; the former case might be relevant for the
deleptonization era, while the latter for the initial cooling era.
It is interesting to see the
difference between
the neutrino-trapped and -free cases:  the curve
is shortened as the entropy increases in the former case,
while elongated in the latter case. These two features are 
essential for
the following argument about the delayed collapse and
maximum mass of the cold neutron stars. 

The delayed collapse is
possible if the initially stable 
star on a curve finds no end point on other curves as a
result of the evolution through deleptonization 
or cooling with the baryon number fixed.
Consider a typical case for example:
A PNS has initially $Y_{le}=0.4$ and $S=2$ 
after supernova explosion
and evolves
through deleptonization to  neutrino-free and $S=2$ stage.
We can clearly see the PNS with 
large enough mass
can exist as a stable star at the beginning but
cannot find any point on the neutrino-free and $S=2$ curve.
Therefore they must collapse to the low-mass black hole
by deleptonization.
It is to be noted that because the $Y_{le}=0.4$ and $S=2$ star
never includes kaon condensate,
its collapse is largely due to the appearance of 
kaon condensate in the core.
Thus
we may conclude that
kaon condensation is very plausible to cause the delayed collapse.

Furtheremore we can determine the maximum mass of 
cold neutron stars by examining the evolution of PNS at 
the initial cooling stage, 
where neutrinos are never trapped. 
Usually we assign the maximum mass of
the cold neutron stars from the mass-radius curve 
by the use of the EOS at $T=0$
( see the $S=0$ and neutrino-free case in Fig.\ref{fig:MR}). 
However, it is wrong 
when we take into account the evolution of neutron stars,
especially in the initial cooling stage\cite{takatsuka}.
As already mentioned, the
entropy dependence of the curve for the neutrino-free case is opposite 
to the neutrino-trapped case. Hence, once a PNS
resides on the large-entropy curve, it necessarily evolves into the
corresponding point on the smaller-entropy curve through initial cooling.
As an example, consider the evolution of a PNS with $S=2$.
Because only the stars with $N_B \le 2.14\times 10^{57}$ at the beginning
can evolve to the neutrino-free and cold ($S=0$) star,
the maximum mass of cold neutron stars can be determined as $1.54 M_\odot$.

In order to study the mechanism of delayed collapse
and mass region which should collapse in more detail,
we had better study the dynamical evolution
beyond the static configurations.
This work is in progress.
As another remaining issue,
we will refine the EOS to include the effects
of thermal kaon loops on the nucleon propagator
and the zero-point energy.

\end{document}